\documentclass[11pt]{article}
\usepackage[english]{babel}
\usepackage[a4paper,top=2cm,bottom=2cm,left=2.2cm,right=2.2cm,marginparwidth=1.75cm]{geometry}
\usepackage{setspace}
\usepackage{amsmath}
\usepackage{graphicx}
\usepackage[colorlinks=true, allcolors=blue]{hyperref}

\usepackage[utf8]{inputenc}
\usepackage[T1]{fontenc}
\usepackage[english]{babel}
\usepackage{amsmath,amssymb}
\usepackage[numbers]{natbib}
\usepackage{hyperref}
\usepackage{titlesec}
\usepackage{caption}
\usepackage{setspace}
\usepackage{multicol}
\usepackage{ragged2e} 
\usepackage[dvipsnames]{xcolor} 
\usepackage{tcolorbox} 
\usepackage{authblk}

\makeatletter

\renewcommand\AB@affilsepx{, }

\def\myinlineaffils{%
  {\small
   \renewcommand{\and}{, }
   \AB@affillist}
}

\renewcommand{\maketitle}{%
  \begin{center}
    {\LARGE\@title\par}
    \vskip 1em
    {\large
      \setlength{\parskip}{0.3em}
      \setlength{\parindent}{0pt}
      \@author
      \par}%
    \vskip 1em
    \vskip 1em
  \end{center}
}
\makeatother

\title{Science Enabled by a 30-Meter-Class Telescope in the Northern Hemisphere: Massive Stars at Low Metallicity}

\author[1]{M. Garcia}
\author[2,3]{A. Herrero}
\author[4]{I. Negueruela}
\author[5]{N. Castro}
\author[6]{S.~R. Berlanas}
\author[1]{M. Cervi\~no}
\author[3,2]{G. Holgado}
\author[7]{J. Iglesias P\'aramo}
\author[7]{C. Kehrig}
\author[6]{J. Ma\'iz Apell\'aniz}
\author[6]{J.~M. Mas-Hesse}
\author[1]{F. Najarro}
\author[3,2]{S. Sim\'on-D\'iaz}
\author[7]{J.~M. Vilchez}

\affil[1]{Centro de Astrobiolog\'ia, CSIC-INTA. Crtra. de Torrej\'on a Ajalvir km 4., E-28850 Torrej\'on de Ardoz (Madrid), Spain}
\affil[2]{Departamento de Astrofísica, Universidad de La Laguna, 38205, La Laguna, Tenerife, Spain}
\affil[3]{Instituto de Astrofísica de Canarias, 38200, La Laguna, Tenerife, Spain} 
\affil[4]{Departamento de Física, Universidad de Alicante, Carretera de San Vicente del Raspeig s/n, E-03690, San Vicente del Raspeig, Alicante, Spain}
\affil[5]{Leibniz-Institut für Astrophysik Potsdam (AIP), An der Sternwarte 16, D-14482 Potsdam, Germany}
\affil[6]{Centro de Astrobiología. CSIC-INTA. Campus of the European Space Astronomy Centre (ESAC), E-28692 Villanueva de la Cañada, Madrid, Spain}
\affil[7]{Instituto de Astrofísica de Andalucía, CSIC, Apartado de correos 3004, 18080 Granada, Spain}

\begin{document}
\maketitle

\newpage

\begin{tcolorbox}[colback=RoyalBlue!5!white,colframe=black!75!black, width=\textwidth]
\justifying
{\noindent Massive stars are at the core of our observations of the Universe up to the reionization epoch, both through their intense ionizing fluxes and through the energetic end products that release fresh elements into the interstellar medium. Our interpretation of very high redshift galaxies and transient phenomena depends on knowledge derived from massive star populations in the Milky Way and nearby galaxies, with characteristics that only remotely resemble the conditions in the early Universe. However, the models supporting these interpretations have been tested in a narrow range of environments and carry significant uncertainties when extrapolated. Advancing in our understanding of the Universe beyond the Local Volume therefore requires extending massive star studies to conditions representative of the early Universe. The next generation of telescopes has the potential to accomplish this goal.}
\end{tcolorbox}

\section{Massive stars at low metallicity}
Understanding the Cosmic Dawn Epoch is among the most exciting challenges of modern Astrophyiscs. How the first stars and galaxies formed, how the reionization of the Universe proceeded, how its chemical composition and Star Formation Rates (SFR) evolved with time, are very compelling open questions. Massive stars, with their very energetic ionizing fluxes processed through the interstellar and intergalactic media are key tools to explore the Universe and interpret the information in the light coming from very early ages, like in the cases of Earendel \citep[at $z=$6.2][]{Welch22} and JADES-GS-z14-0 \citep[at $z=$14.32][]{Carniani24}. Moreover, their final products like Supernovae, neutron stars, Gamma-Ray Bursts, black holes and gravitational wave events (GWE) trace the high energy Universe.

Using massive stars as cosmic tools requires an understanding of the physics governing their properties and evolution. Yet, physical conditions vary greatly across space and time. On the one hand, SFR increases, peaks at redshift $z \sim$2 and then decreases with time. On the other, the abundances of chemical elements heavier than helium increased by orders of magnitude since the Big Bang until the present day \citep{Madau14}. 



Metallicity (Z) is a fundamental parameter for the stellar structure, but in the case of massive stars it also determines the strength of their stellar winds, affecting not only the mass lost by the star during its evolution, but also the opacity of their external layers and the spectral distribution of its ionizing flux. Their mechanical and ionizing feedback in turn strongly alters the properties of the surrounding medium \citep{Kudritzki00, Puls08, Josiek24}. 

Other physical processes also play an important role in modifying the structure and evolution of massive stars. A large fraction of them are born in binary or multiple systems, opening new evolutionary channels through binary interaction and mergers \citep{Sana12, Mink13, Marchant24}. Rotation drives mixing processes bringing nuclear burning products and angular momentum to the surface, whilst magnetic fields, pulsations and turbulence alter the structure of the stars and their final fate \citep{Langer12}. These effects are not independent of each other, and for instance binary interaction modifies the rotational properties of the stars through tides and mass and angular momentum transfer. Underlying all these processes, the stellar chemical composition shapes them to a significant extent, affecting star formation and evolution across the Universe.

There have been important advances in our understanding of massive stars. 
The Milky Way (MW) and the Magellanic Clouds (MCs) have enabled the exploration of the 2 to 1/5 Z$_\odot$ range. We have characterized stellar winds present in massive stars down to Z$_{\rm SMC}$ \citep[e.g.,][]{Mokiem07, Puls08}, developing a theory that can potentially be extended towards lower metallicities.
However, uncertainties persist or have newly emerged, when 
approaching the average chemical composition of early epochs. It is unclear whether the currently accepted Radiatively Driven Wind (RDW) theory breaks for (relatively) low luminosities as well as for low metallicities \citep[e.g.,][]{Bouret03, Garcia25}. This may have a significant impact on our estimated budget of ionizing photons by reducing the opacity in the wind and photosphere. 

Fundamental predictions concerning the behaviour of massive stars at low Z are presently being challenged. Contrary to expectations, there is no  evidence that stars at low metallicities undergo chemically homogeneous evolution, a hypothetical channel that may result in very different evolutionary paths, power He{\sc II} nebular emission and lead to the formation of massive double black holes in close orbits and gravitational waves \citep{Szecsi15, Mink16}.
It remains also unclear whether the upper mass limit increases with decreasing Z, a key property for the expectation of extreme masses for the first stars ($\sim$ 1\,000 M$_\odot$, \cite{Klessen23}), and a consequence of inhibited gas fragmentation in the absence of enough metals to cool down molecular gas. There is intriguing evidence that a substantial fraction of the black hole masses inferred from GWE are significantly higher than those of the most massive black holes measured in the Local Universe \citep{Abac25}. Since the winds of massive stars are known to weaken markedly with decreasing Z \citep[e.g.,][]{Garcia25}, current observations strongly suggest that stellar physics changes once a critical Z threshold is reached. Finally, the evolution of low-metallicity stars with masses above $\approx$70 M$_{\odot}$ may lead to a pair-instability supernova \citep{Renzo26} that entirely disrupts the star, leaving no remnant. But at even higher masses or lower metallicities, more massive CO cores may instead undergo photodestintegration instability and directly collapse into high-mass black holes. These events are believed to manifest as superluminous supernovae \citep[e.g.,][]{Moriya26} or other exotic transients whose extreme luminosities may render them powerful tracers of the early universe. 

These uncertainties amplify when we try to extend our knowledge below Z$_{SMC}$. The Initial Mass Function (IMF), chemical yields, binary fractions, mechanical feedback, fluxes ionizing the surrounding medium and stellar end products are strongly affected, while critical for synthetic stellar populations and the interpretation of observations when no spatial resolution is possible \citep{Leitherer14, Eldridge17}. 

\section{Our laboratory: the nearby low metallicity star-forming galaxies} 
The lucky circumstance of having the MCs close to our Galaxy has allowed dedicated surveys to explore the Z-effects in the range covered by all three galaxies \citep{Evans05, Evans11, Simon20}. However, to extend our knowledge beyond this range we need to expand our observations to farther galaxies.

\begin{figure}[!h] 
	\resizebox{\hsize}{!}{\includegraphics[angle=0,width=0.1\textwidth]{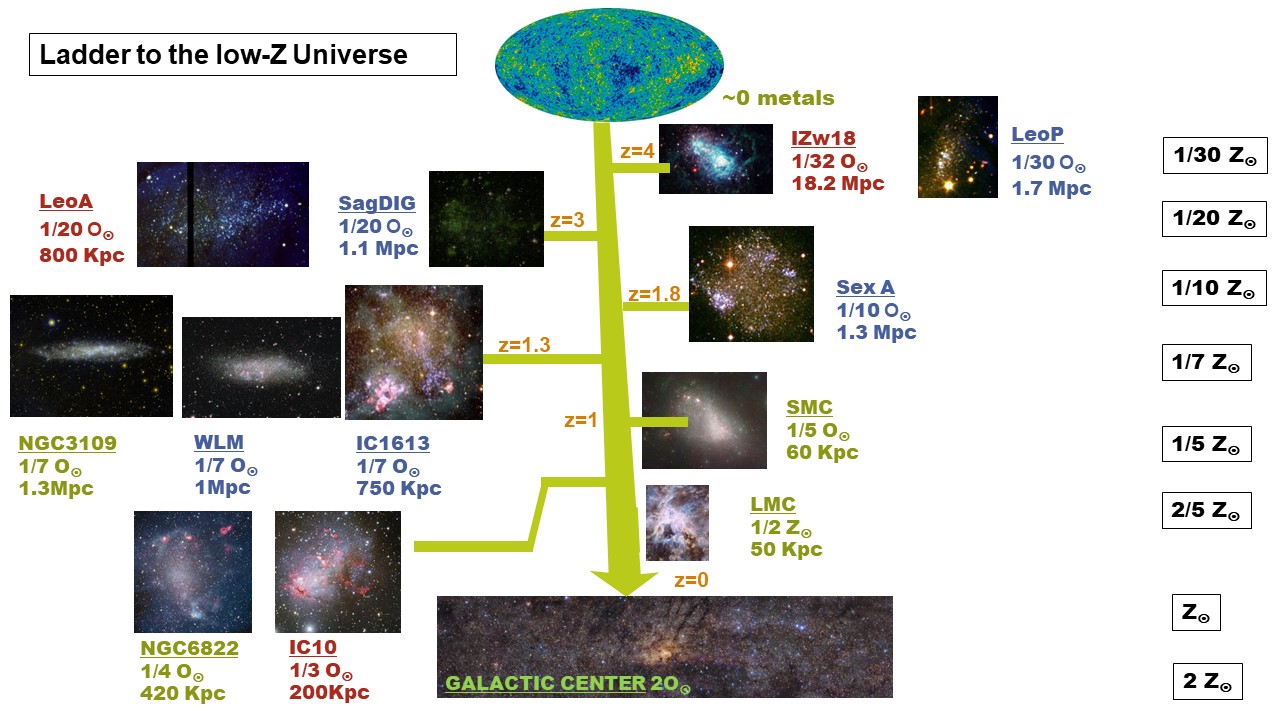}}
	\caption{Metallicity ladder to Cosmic Dawn. Colours represent the sky hemisphere where galaxies can be observed: North (red), South (green) or both (blue). Present-day facilities allow us to reach 1/10 Z$_\odot$ metallicities. Beyond, larger telescopes of the 30-m class are required. I Zw18, with very low-Z and strong SFR, is the ideal environment to confront massive star evolution and RDW theories.}
	\label{Fig:ladder} 
\end{figure}

Fig.~\ref{Fig:ladder} represents the way in which this objective could be reached. It shows different galaxies labeled by their distance, representative Z and redshift, and coloured according to their declination. The MW and the MCs constitute the first steps in this ladder. But with a characteristic Z between 0.1 and 0.01 Z$_\odot$ at the peak of SFR in the Universe, their 
metallicity is way too high. 

Sex~A breaks the 0.1 Z$_\odot$ frontier required to enter the deep Universe, and simultaneously represents the distance limit for extensive spectroscopic surveys of individual massive stars with present-day instrumentation. The recent catalog of massive stars in Sex A by \cite{Lorenzo22} provides a solid ground for extending our Z baseline down to values representative of the young Universe. However, it also constitutes a clear example of the challenge we face in this endevour. 

Beyond Z$_{\rm SexA}$ (and at larger distances) we find some galaxies with a strong potential for studies at low metallicity. The challenge to enter into the typical Z range of the early epochs
means reaching galaxies like Leo A, Leo P, SagDig and specially I Zw18 (that combines very low Z with high SFR).

Exploiting the potential offered by the low-Z nearby galaxies implies to carry out large spectroscopic surveys. At the distances involved, the requirement of high signal-to-noise ratio and spectral and spatial resolution is highly demanding. At the same time, those surveys have to be very extensive, to allow a statistically significant exploration of all possible massive stars evolutionary channels. 

\section{The crucial role of the Northern Hemisphere: IC~10 and I~Zw18}
The debate whether the upper-end of the IMF depends on Z cannot be resolved within the Local Group. The most massive resolved stars known to date have $\gtrsim$200M$_\odot$ and are located in the Large Magellanic Cloud (LMC, 1/2 Z$_\odot$) at the core of the 30~Doradus star forming region \citep{Crowther10, kalari22}. World-wide efforts to scrutinize metal-poorer environments have failed to detect more massive stars than 60-80 M$_\odot$ \cite{Lorenzo25, Bestenlehner25}. Deep infrared photometric studies pending, evidence suggests that most nearby metal-poor galaxies have too low present-day SFR and gas reservoir to form stars over 100 M$_\odot$. The two most promising candidates for further progress are located in the northern hemisphere.

With lower metallicity than the LMC and sufficiently high SFR, IC~10 is the only other metal-poor galaxy of the Local Group that could host very massive stars (VMSs) and enable studies of the metallicity dependence of the IMF. However, this galaxy has been scarcely studied because it experiences high internal extinction, has a low galactic latitude (-3$^{\circ}$) and blue massive stars are faint in the IR. \cite{Crowther03} identified 26 WR stars, but observations squeezed the Gemini-N telescope to their limits and deeper surveys will require a 30m-class telescope. 

Nonetheless, the most important landmark is I Zw18, that exhibits a present-day SFR= 0.6 M$_\odot$ yr$^{-1}$ \citep{Bortolini24}. With a rich gas reservoir, I~Zw18 foreseeably hosts a wealthy population of massive stars that makes it the ideal place to figure out low-Z specific evolutionary pathways and look for VMSs. It is also key to understand the origin of He{\sc II} and other high ionization nebular emission lines, whose origin remains a mistery and keeps challenging current stellar models in the low Z regime \citep{Eldridge22, Kehrig21, Szecsi25}. The relevance of this galaxy is such that it is one of the highlighted scientific cases of the Habitable Worlds Observatory (HWO), and resolving massive stars in this galaxy is a chief driver for one of the proposed instruments: an integral field unit operating in the UV \citep{Senchyna25}. 
However, analog observations attempting to resolve its population in the NIR and optical ranges, critical to determine stellar properties,  will never be carried out with the current plans for instrumentation. The task is unfeasible with 10m-class telescopes, the JWST is not sensitivity enough, and the galaxy is out of reach to the ELT.

So far, our observations in Sex A have only begun to probe early-Universe conditions. The next step in the Z ladder (Fig.~\ref{Fig:ladder}) will become accessible with 30 m-class telescopes equipped with high-multiplex spectrographs and adaptive optics. Combined with space-based facilities such as HST, JWST, and HWO, these capabilities will enable us to study the massive-star populations of galaxies beyond Sex A. Ultimately, our goal is to characterize massive stars and their interaction with the interstellar medium in a galaxy like I Zw18, either through observations of individual stars or population synthesis analyses of small clusters, but grounded in robust knowledge of low-Z stars.

This major effort will provide a solid basis for interpreting observations across many areas of astrophysics, from extremely metal-poor stars and high-redshift galaxies to gamma-ray and gravitational wave progenitors, from the peak of cosmic star formation to the epoch of reionization.




\begin{multicols}{2}
\renewcommand{\bibfont}{\footnotesize}  
\bibliographystyle{unsrtnat}     
\bibliography{refs_lowZ}
\end{multicols}


\end{document}